\documentclass[10pt]{iopart}
\usepackage{iopams}  
\usepackage{IEEEtrantools}
\usepackage{amsfonts} 
\usepackage{graphicx} 

\begin{document}
\title{Evolution and invariants of free-particle moments}
\author{Mark Andrews}
\ead{mark.andrews@anu.edu.au}    
\address{Department of Quantum Science, Research School of Physics \\Australian National University,  Canberra ACT 0200 Australia}

\begin{abstract}
Moments are expectation values of products of powers of position and momentum, taken over quantum states (or averages over a set of classical particles).  For free particles, the evolution in the quantum case is closely related to that of a set of classical particles. Here we consider the evolution of symmetrized moments for free particles in one dimension, first examining the geometric properties of the evolution for moments up to the fourth order, as determined by their extrema and inflections. These properties are specified by combinations of the moments that are {\it invariant} in that they remain constant under free evolution. An inequality constrains the fourth-order moments and shows that some geometric types of evolution are possible for a quantum particle but not possible classically, and some examples are examined. Explicit expressions are found for the moments of any order in terms of their initial values, for the invariant combinations, and for the moments in terms of these invariants.
\end{abstract} 

\maketitle

\section{Introduction}
Moments of quantum states (or of a set of classical particles) are commonly used to give simple averaged properties of position or momentum (or of their powers and products).  In quantum mechanics the first-order moments ($n=1$) are the expectation values of the position operator $\hat x$ or the momentum operator $\hat p$. Thus $\langle \hat x\rangle$ and $\langle \hat p\rangle$ define the centroid of the state, and according to Ehrenfest's theorem the centroid of a free particle follows a classical evolution; that is, it has a constant velocity. The higher-order moments will always be taken to be relative to the centroid. Thus the second-order moments include  $\langle(\hat x-\langle \hat x\rangle)^2\rangle$,  $\langle(\hat x-\langle \hat x\rangle)(\hat p-\langle \hat p\rangle)\rangle$, and  $\langle(\hat p-\langle \hat p\rangle)^2\rangle$.

The second-order moment  $\langle(\hat x-\langle \hat x\rangle)^2\rangle$ is widely used to provide a measure of the spatial spread of the state, usually expressed as $\sigma_2=\langle(\hat x-\langle \hat x\rangle)^2\rangle^{1/2}$. Its evolution is simple: it is symmetric in time about a minimum waist. The second moment of momentum, $\langle\hat p^2\rangle$ is proportional to the energy, while $\langle(\hat p-\langle \hat p\rangle)^2\rangle$ gives a measure of the spread in momentum and combines with the spread in position to give Heisenberg's uncertainty relation.

The third-order moment $\langle(\hat x-\langle \hat x\rangle)^3\rangle$ gives a measure of the skewness of the state. This moment can be positive or negative, and usually will change sign during its evolution, which may have local extrema. The most appropriate length to represent the skewness is $\mathcal{S}=\langle(\hat x-\langle \hat x\rangle)^3\rangle / \langle(\hat x-\langle \hat x\rangle)^2\rangle$.

The fourth-order moment $\langle(\hat x-\langle \hat x\rangle)^4\rangle$ leads to the length $\sigma_4=\langle(\hat x-\langle \hat x\rangle)^4\rangle^{1/4}$ that gives greater weight to the outer parts of the state than does $\sigma_2$; but we will see that its evolution can be more complex, and may have a local maximum. A dimensionless measure comparing the two spreads is the kurtosis, $K=\langle(\hat x-\langle \hat x\rangle)^4\rangle/\langle(\hat x-\langle \hat x\rangle)^2\rangle^2$. For example, the Gaussian wavefunction $\psi(x)=\exp(-x^2/a^2)$ has $\sigma_2=\frac{1}{2}a$ and $K=3$. Examples with longer tails are $\exp(-|x|/a)$ with $\sigma_2\approx 0.707\,a,\,K=6$, and $\exp(-\sqrt{|x|/a})$ with $\sigma_2\approx 2.74\,a,\,K=25.2$.

We will show that the asymptotic behaviour of any moment of order $n$ for large times (past or future) is determined by the moment $\langle(\hat p-\langle \hat p\rangle)^n\rangle$.

The operators $\hat x$ and $\hat p$ do not commute and we will consider only \textit{symmetrized} moments. The symmetrized moments of order $n$ are expectation values averaged over all possible orderings of products with a fixed number $k$ of occurrences of $\hat p-\langle \hat p\rangle$ and $n-k$ occurrences of $\hat x-\langle \hat x\rangle$ with $k=0, 1, 2, ...n$, and we will see that the symmetrized moments of order $n$ form a closed set as they freely evolve.

Here we consider in detail the evolution of these symmetrized moments. In particular, we will find explicit expressions for the moments in terms of their initial values, and that these expressions are polynomial in the time. Then we give greater attention to the third and fourth order moments, particularly with regard to the extrema and inflections of the moments as they evolve. We find that the essential information about these critical points in the evolution is expressed in certain combinations of the moments that are invariant in the sense that they do not change as the wavefunction freely evolves. This allows us to predict the general features of the evolution of these averaged aspects of wave packets, in terms of their initial values. Expressions are given for these invariants for moments of any order.

Some exotic aspects of the evolution are purely quantum. For example, local maxima in the evolution of $\langle(\hat x-\langle \hat x\rangle)^4\rangle$ can occur for some wavefunctions but never for a set of classical particles. We derive an inequality that constrains this behaviour.

Some basic aspects of moments and their applications were discussed in \cite{Bal}\,-\,\cite{BrizGen2}, and the observability of non-classical features in \cite{Vogel}.

\section{Dynamics of a free particle}
A free quantum particle has the Hamiltonian $\hat H=\hat p^2/2m$. We use the notation \cite{A} 
\begin{equation} \label{eqn:DA}
\textrm{D}_t \hat{A}:=\frac{\partial \hat{A}}{\partial t}+\frac{\imath}{\hbar}[\hat{H},\hat{A}].
\end{equation}
$\textrm{D}_t \hat{A}$ is the `total time-derivative' of the operator $\hat{A}$, which adds to the partial derivative a term that takes account of the evolution of the state under the Hamiltonian $\hat H$ in such a way that
$\textrm{d}_t\langle\hat{A}\rangle=\langle \textrm{D}_t \hat{A}\rangle$ for any state and any operator $\hat{A}$, where $\textrm{d}_t$ stands for the usual time-derivative $d/dt$. Also $\textrm{D}_t (\hat{A}\hat{B})=(\textrm{D}_t \hat{A})\hat{B}+\hat{A}(\textrm{D}_t \hat{B})$, similar to ordinary differentiation. (These equations are similar to those for the Heisenberg picture\cite{HP}, but here we retain the Schr\"odinger picture where the states change with time, and the operator $\hat x$ does not change.)

For the free Hamiltonian $\hat H=\frac{1}{2m}\hat p^2$, and
\begin{equation}\label{eqn:motion}
\textrm{D}_t \hat{x}=\hat{p}/m,  \qquad \textrm{D}_t \hat{p}=0,
\end{equation}
similar to the equations, $\textrm{d}_t x=p/m$ and $\textrm{d}_t p=0$, for the classical particle.

\subsubsection*{First Moments -- the motion of the centroid.} From \eref{eqn:motion}, 
$\textrm{d}_t \langle\hat{x}\rangle=\langle\hat{p}\rangle/m$, $\textrm{d}_t \langle\hat{p}\rangle=0$, the same as the usual classical equations, with the solution $\langle\hat{x}\rangle=x_0+\langle\hat{p}\rangle t/m$ where $\langle\hat{p}\rangle$ is constant and $x_0$ is the initial position of the centroid.


\subsubsection*{Moments relative to the centroid.}
It is convenient to introduce $\hat X :=\hat x - \langle\hat x\rangle$ and $\hat P :=\hat p - \langle\hat p\rangle$, and then
\begin{equation}\label{eqn:motionRel}
\textrm{D}_t \hat{X}=\hat{P}/m,  \qquad \textrm{D}_t \hat{P}=0.
\end{equation}

\subsubsection*{Second Moments.} For a particle in one dimension, there are three symmetrized second moments:
\begin{equation}
Y_0 =\langle \hat X^2\rangle,\qquad
Y_1=\case{1}{2}\langle\hat X\hat P+\hat P\hat X\rangle,\qquad
Y_2 =\langle \hat P^2\rangle.
\end{equation}
The moment $Y_0$ gives a measure of the spatial spread of the wavefunction via $\sigma_2=Y_0^{1/2}$, and $Y_1$ measures the correlation between position and momentum. The moment $Y_2$, in addition to giving a measure of the spread in momentum, is related to the positive \emph{quantal-energy} $\epsilon=Y_2/2m=\langle\hat H\rangle\,-\,E_c$, where $E_c=\langle\hat p\rangle^2/2m$ is the classical energy of the centroid. Both $\langle\hat H\rangle$ and $E_c$ remain constant. It is a simple matter to find the time-derivatives of these moments from \eref{eqn:motionRel}:
\begin{equation}
\textrm{d}_t Y_0=2Y_1/m,\qquad \textrm{d}_t Y_1=Y_2/m,\qquad\textrm{d}_t Y_2=0,
\end{equation}
and the evolution of the moments in terms of their initial values $y_0, y_1, y_2$ is
\begin{equation}\label{ev2}
Y_0=y_2 t^2/m^2+2y_1t/m+y_0,\;\;\;  
Y_1=y_2t/m+y_1,\;\;\; 
Y_2=y_2. 
\end{equation}
Along with the energy constant $Y_2$, the combination
\begin{equation}\label{eqn:Z1}
 \Omega_2= Y_0 Y_2-Y_1^2
 \end{equation}
 is important because it is constant, $\textrm{d}_t\Omega_2=0$, and subject to the inequality $\Omega_2\geq \frac{1}{4}\hbar^2$, which is stronger than Heisenberg's uncertainty relation $Y_0 Y_2\geq \frac{1}{4}\hbar^2$.  [The stronger inequality was originally proved by Schr\"odinger in 1930. It is easily derived from Schwarz's inequality $\langle \hat P^2\rangle \langle \hat X^2\rangle \geq |\langle \hat P\hat X\rangle|^2$ using $\hat P\hat X=\frac{1}{2}(\hat P\hat X+\hat X\hat P -\imath\hbar)$.]

\Eref{ev2} show that $Y_0$ is quadratic in $t$ with the minimum value $\Omega_2/y_2$ at the time $t_0=-m\,y_1/y_2$, the time when the correlation $Y_1$ is zero. That is, the `uncertainty product' $Y_0 Y_2$ takes its minimum value $\Omega_2$ at just one time $t_0$.

We will refer to quantities, like $Y_2$ and $\Omega_2$ for the second-order moments, as \textit{invariants}: they remain constant in time through the equations of motion.

\subsubsection*{Moments over a set of free classical particles.} If the $\mu$th particle has position $x_\mu$ and momentum $p_\mu$, then the equations of motion are $\textrm{d}_t x_\mu =p_\mu/m$ and $\textrm{d}_t p_\mu =0$. The centroid has position $\bar x=N^{-1}\Sigma_\mu x_\mu$ and momentum $\bar p=N^{-1}\Sigma_\mu p_\mu$, where $N$ is the number of particles. Also $\bar x$ and $\bar p$ satisfy the same equations of motion. (We are assuming, for simplicity, that the particles are all of the same mass; but the results can be easily extended to cover different masses.) Then the deviations from the centroid, $X_\mu=x_\mu-\bar x$ and $P_\mu=p_\mu-\bar p$, also satisfy the same equations of motion: $\textrm{d}_t X_\mu =P_\mu/m$ and $\textrm{d}_t P_\mu =0$.

The moments of order $n$ about the centroid are $\mathcal{Y}_k=N^{-1}\Sigma_\mu P^k X^{n-k}$ and it follows from \eref{eqn:motionRel} that
\begin{equation}\label{eqn:dYclass}
\textrm{d}_t\mathcal{Y}_k=(n-k)\mathcal{Y}_{k+1}\,/m.
\end{equation}
The set $\mathcal{Y}_k$ with $k=1, 2, ... n$ is closed under evolution in the sense that the set of values at any time determine the values at any later time.

\section{Higher order Moments}
Our main purpose is to similarly examine the evolution and invariants of symmetrized moments of order $n>2$. For any order $n$, the symmetrized quantum moment $Y_k$ is the expectation value averaged over all products that contain $\hat X$ exactly $n-k$ times and $\hat P$ exactly $k$ times. The index $k$ ranges from 0 to $n$. For example, with $n=3$,
\begin{equation}\label{eqn:Y3def}\eqalign{
Y_0= \langle \hat X^3\rangle,\\
Y_1= \case{1}{3}\langle \hat X^2\hat P+\hat X\hat P\hat X+\hat P\hat X^2\rangle,\\
Y_2 = \case{1}{3}\langle \hat P^2\hat X+\hat P\hat X\hat P+\hat X\hat P^2\rangle,\\
Y_3= \langle \hat P^3\rangle.}
\end{equation}
Note that, to keep the notation simple, we use the same symbol $Y_k$ for the moments for all values of $n$, even though they will differ for each $n$.

Any moment of order $n$ can be expressed, using $[\hat P,\hat X]=-\rmi\hbar$, in terms of the set of symmetrized moments of order $n$.\cite{AH} It will now be shown that these symmetrized moments have the same evolution equations as the moments of a set of classical particles, and the same set of invariants.

\subsubsection*{Quantum evolution of symmetrized moments.} In the calculation of the time-derivative of $Y_k$ no commuting of operators is involved, because the result is always a symmetrized moment, and must be the same as in \eref{eqn:dYclass}. Therefore
\begin{equation}\label{eqn:dtY}
\textrm{d}_t Y_k=(n-k)Y_{k+1}/m,
\end{equation}
where $k=0,1,2,...,n-1$. It follows that the moments $Y_k$ at time $t$ in terms of the initial values $y_k$ is (equation (22) of \cite{BrizGen2})
\begin{equation}\label{eqn:Yy}
Y_k=\sum\nolimits_{\ell=0}^{n-k}\,(^{n-k}_{\;\;\ell})\,(t/m)^{\ell}\,y_{k+\ell}.
\end{equation}
[This is easily proved by taking the time-derivative and replacing $\ell$ by $\ell-1$.]\\
For example, with $n=3$,
\begin{equation}\label{eqn:Matrix3}
 \left[ \begin{array}{c}
Y_0 \\ Y_1\\ Y_2 \\ Y_3 
\end{array}\right]
=	
\left[ \begin{array}{cccc}
1 & 3t/m & 3(t/m)^2 & (t/m)^3\\
0 & 1 & 2t/m & (t/m)^2\\
0 & 0 & 1 & t/m\\
0 & 0 & 0 & 1
\end{array}\right]
\left[ \begin{array}{c}
y_0 \\ y_1\\ y_2 \\ y_3 
\end{array}\right].
\end{equation}
It follows, for all $n$, that $Y_k\sim (t/m)^{n-k}y_n$ as $|t|\to\infty$.

\subsection{Some general features of moments of any order.}
\hspace{7mm} All symmetrized moments are real. (They can all be expressed as the expectation value of an Hermitian operator.)

Spatially symmetric (or antisymmetric) wavefunctions will remain symmetric (or antisymmetric) as they evolve and all moments of odd order will be zero. In particular, such wavefunctions will have no skewness.

In any case, $Y_n$ will be invariant because the operator $\hat P^n$ commutes with the free Hamiltonian.

For even order, any (nonzero) wavefunction, both $Y_0$ and $Y_n$ will be positive and nonzero. There is also a generalisation of the usual uncertainty relation for $n=2$ that has the form $Y_0\,Y_n\geq c_n\hbar ^n$, where $c_n$ is a positive constant (equation (57) of \cite{BrizGen}). In particular, $c_2=\case{1}{4}$ (the Heisenberg uncertainty relation) and $c_4=\case{3}{8}$.

For odd order, it is possible to have $y_n=0$. (As will shortly be shown, this will occur when the initial wavefunction is real.) Then $Y_n=0$ and $\textrm{d}_t Y_{n-1}=Y_n/m$ implies that $Y_{n-1}$ is constant. Then the analysis of the evolution follows in a similar way to that for order $n-1$, although the numerical coefficients are different. 

There are also implications for initially real wavefunctions, often used in illustrative examples. We ignore any phase factor that is independent of position -- it would not effect the moments. (Note that any initially real wavefunction will immediately become complex as it freely evolves.) Any Hermitian Hamiltonian will have a basis of eigenfunctions that can be taken to be initially real. 

If the initial wavefunction is real, all moments $Y_k$ will be initially zero if $k$ is odd (because there is an odd number of momentum operators and each has a factor $\imath$, but the moment must be real). This implies that the evolution of $Y_0$ can be expressed in terms of powers of $t^2$ and its analysis becomes much simpler. For example, if $n=4$, then $Y_0$ is quadratic in $t^2$.

\section{Invariants and generic shape of moment evolution}

The general shape of the evolution of the moments follows from their maxima, minima, and inflections, and we will see that these critical points in the evolution are directly related to invariant combinations of the moments. Their invariance is explained by the fact that local features of the evolution (such as the values of the moments at extrema and inflections) are independent of the time-origin. 

From \eref{eqn:dtY} the extrema of $Y_k$ occur at the zeros of $Y_{k+1}$ and the inflections of $Y_k$ occur at the zeros of $Y_{k+2}$. Taking the times to be measured relative to $t_0=-m\,y_{n-1}/y_n$ will reveal that the existence of the critical points (and therefore the shape of the evolution) is directly related to the invariants

In this way, we will analyse the evolution of moments of the third and fourth order in the following two sections and in \sref{section:Z} the form of these invariants will be extended to arbitrary order $n$.

\section{Third-order moments -- skewness.} \label{section:order3}
The moment $Y_0=\langle X^3\rangle$ gives a measure of the skewness of the wavefunction; it can be positive, zero, or negative.

\subsubsection*{The standard case, with $y_3\neq 0$.}  With $n=3$, \eref{eqn:dtY} gives $Y_2=y_2+y_3t/m$. It is convenient to express the time-dependence of the moments in terms of
\begin{equation}
	u=(t-t_0)/m,
\end{equation}
where $t_0=-m\,y_2/y_3$; then $Y_2=y_3\,u$. From \eref{eqn:Matrix3},
\begin{equation}
Y_0=y_0+3y_1t/m+3y_2(t/m)^2+y_3(t/m)^3,	\label{eqn:Yy3}
\end{equation}
 and after substituting $t\to m u+t_0$,
\begin{equation}\label{eqn:Y0y3}
	Y_0=y_3\,u^3 +3(y_1y_3-y_2^2)\,u/y_3+
	(y_0 y_3^2-3y_1 y_2 y_3+2y_2^3)/y_3^2.
\end{equation}
It can be easily verified by direct differentiation that
\begin{equation}\label{eqn:Z2}
	\Omega_3 = Y_1Y_3-Y_2^2 \;\;\;\;\textrm{and}\;\;\;
	\Lambda_3 = Y_0 Y_3^2 - 3\,Y_1 Y_2 Y_3 +2\,Y_2^3 
\end{equation}
are invariants; they relate to features unchanging under time-translation. 

\vspace{2mm}
\Eref{eqn:Y0y3} leads to
\begin{equation}\label{eqn:3YT}\eqalign{
	Y_0=y_3\,u^3 + 3(\Omega_3/y_3)u + \Lambda_3/y_3^2,\\
	Y_1=y_3\,u^2 + \Omega_3/y_3,\\
	Y_2 =y_3\,u ,}
\end{equation}
where $Y_1$ is obtained by differentiating $Y_0$, and similarly for $Y_2$ from $Y_1$. The invariant quantities $y_3, \Omega_3$, and $\Lambda_3$ emerge from the values of the moments at $u=0$. 

From these expressions for the moments in terms of the invariants, which can be easily calculated from the initial moments, we can deduce the general behaviour of the time evolution of the moments. 

The moment $Y_1$, which is quadratic in $u$, takes its minimum value of $\Omega_3/y_3$ at $u=0$, and if $\Omega_3 \geq 0$ then $Y_1$ cannot be zero and  hence $Y_0$ will have no extrema but will have an inflection at $u=0$. Otherwise, if $\Omega_3 < 0$ then $Y_1$ will be zero at $u=\pm u_0$ where $u_0 =\sqrt{|\Omega_3|}/y_3$. It follows that $Y_0$ will take its maximum of $Y_0^+$ and its minimum of $Y_0^-$ where
\begin{equation}
	Y_0^{\pm}=(\Lambda_3 \pm 2\,|\Omega_3|^{3/2} )/y_3^2.
\end{equation}

\begin{figure}\label{fig:Generic3}
\centering
\includegraphics[width=85mm]{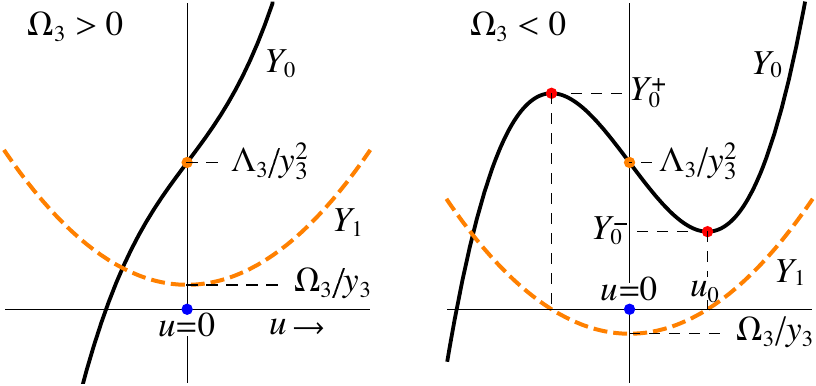}
\caption{Generic forms for third moments $Y_0$ and $Y_1$ (when $y_3>0$).} 
\end{figure}

Thus, as illustrated in figure 1, the general properties of the skewness $Y_0$ of any wave packet is determined by the values of the invariants $y_3$, $\Omega_3$, and $\Lambda_3$, Also, $t_0$ gives the reference point $u=0$ for the time.

\subsection{In summary: how the skewness $Y_0$ evolves.} We assume here that $y_3 > 0$. [If $y_3 < 0$, the evolution will be the same as if the time (and $y_1$) are reversed.] There are two cases:

$\Omega_3\geq 0$. This is the simplest case (and the other case seems to be difficult to achieve). As time increases, the skewness $Y_0$ never decreases; but it has a point of inflection at the time when $u=0$, where it takes the value $\Lambda_3/y_3$.

$\Omega_3 < 0$. For initial times much earlier than $-u_0$, the skewness $Y_0$ increases to a maximum at time $-u_0$, where it takes the value $Y_0^+$. Then $Y_0$ decreases, through an inflection at time $u=0$ where it takes the value $\Lambda_3/y_3$, and on to a minimum at time $u_0$, where it takes the value $Y_0^-$. After this minimum, $Y_0$ increases uniformly to its asymptotic value $Y_0 \sim y_3\,u^3$ as $t \to \infty$.

\subsubsection*{The special case where $y_3=0$.} \Eref{eqn:Yy3} shows that $Y_3= 0$ and $Y_2$ is constant. Then $Y_1=y_1 + 2y_2\,t/m$ and
$Y_0=y_0 + 3y_1\,t/m+3y_2\,(t/m)^2$. These equations are similar to the general case with $n=2$: one can take $t_0=-my_1/2y_2$ (assuming $y_2\neq 0$) and $u=(t-t_0)/m$, and proceed in the same way.

All wavefunctions that are initially real will have $y_3=0$ and also $y_1=0$, which implies $t_0=0$ and the extremum of $Y_0$ will occur at $t=0$.

\subsection{Example of the evolution of third moments.} A symmetric or anti-symmetric wavefunction will never become skew. An initially real wavefunction will have $y_1=y_3=0$\,; it may acquire skewness as it evolves (if $y_2\neq0$) but the evolution will be quadratic in the time, which is not typical. We will take a simple example that is asymmetric and complex: the wavefunction
\begin{equation}\label{eqn:skewfn}
	\psi(x)=(1+\imath\,b\,x/a)\,\exp(-\case{1}{2}\,x^2/a^2)
\end{equation} 
has $y_0=y_2=0$ and  therefore $\Lambda_3=0$ and the skewness $Y_0$ is $y_3 u^3 +3\Omega_3 u/y_3$. Also $t_0=0$ and $u=t/m$. This wavefunction is not skew initially, but develops skewness as it evolves. The evolution of $Y_0$ is as in figure 1, with $\Lambda_3=0$. Although $\langle x \rangle=0$ initially, $\langle p \rangle=2 b(b^2+2)^{-1}\hbar/a$; so the centroid moves with constant speed.

If $\Omega_3<0$ the skewness will have a maximum of $Y_0^+=2\,|\Omega_3|^{3/2}/y_3^2$ at time $u=-u_0$ where $u_0=\sqrt{|\Omega_3|}/y_3$, an inflection (with $Y_0=0$) at $u=0$, and a minimum of $-Y_0^+$ at time $u=u_0$. If $\Omega_3<0$ there will be no extrema. Calculation gives
\begin{equation}
\!\!\!\!\!\!y_1=-\frac{2 b^3a\hbar}{(2+b^2)^{2}},\;\;\;  y_3=\frac{2 b^3(2-3b^2)\hbar^3}{(2+b^2)^{3}a^3},\;\;\; 
\Omega_3= \frac{4 b^6 (3 b^2-2)\hbar^4}{(b^2+2)^{5}\,a^2}.
\end{equation}
 Then $\Omega_3$ is negative if $b^2<2/3$ and the minimum value of $\Omega_3$ is $\approx -0.0027\,\hbar^4/a^2$ when $b\approx \pm 0.671$. With $b=0.671$, $y_3\approx\,0.027\,\hbar^3/a^3$ and $u_0\approx 1.94\,a^2/\hbar$.  The exact evolution of $\psi(x)$ is easily calculated and is illustrated in figure 2. To compare the skewness with the spread of the wavefunction, we use the skewness-length  $\mathcal{S}=\langle X^3\rangle/\langle X^2\rangle$. At time $u=a^2/\hbar,\;\sigma_2\approx 1.03\,a$ and $\mathcal{S}\approx-0.26\,a$.

\begin{figure}\label{fig:SkewArray}
\includegraphics[width=130mm]{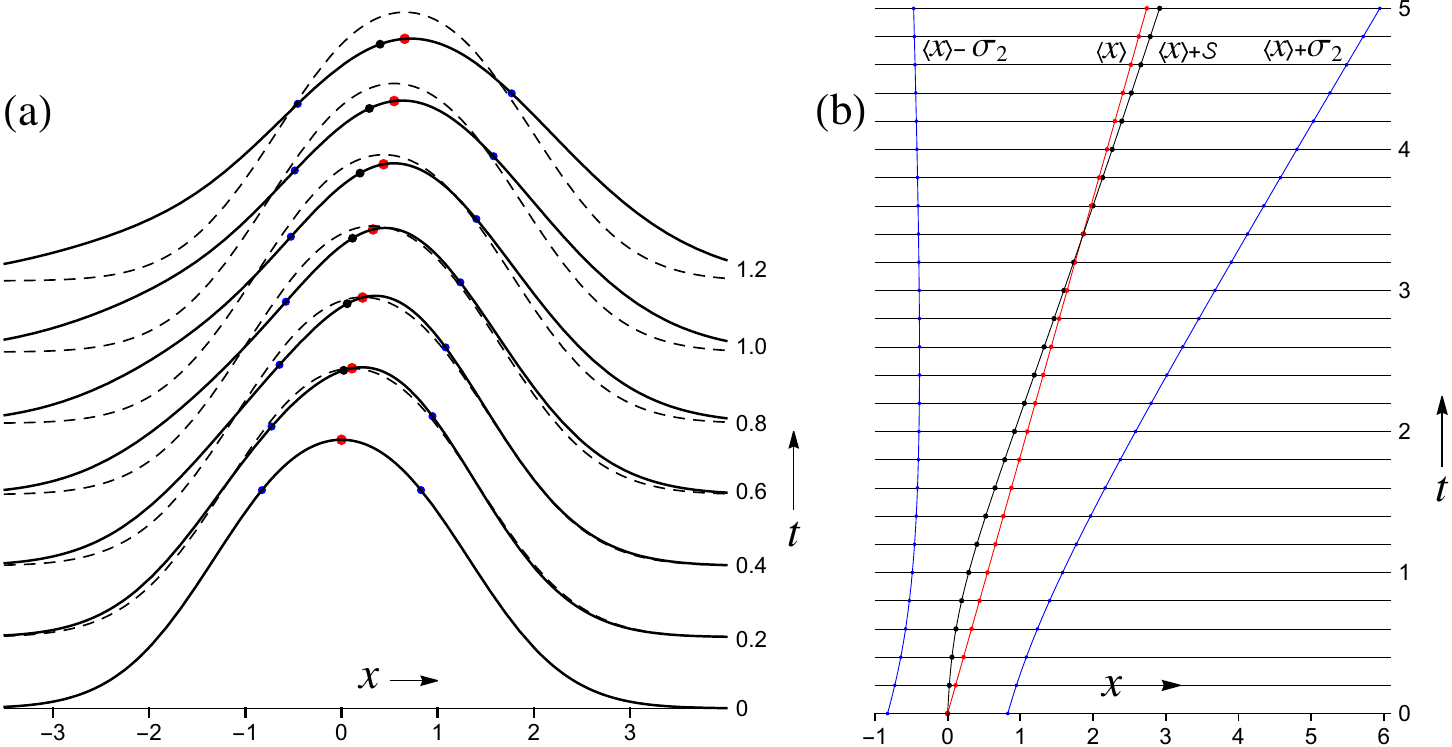}
\caption{Evolution of the wavefunction $\psi(x)$ in \eref{eqn:skewfn} with $b=0.671$. \\ 
(a) Shows $|\psi(x)|$ at a sequence of times increasing by 0.2 (in units of $ma^2/\hbar$) with the distance $x$ in units of $a$. Each successive curve is raised by a fixed amount. The row of larger (red) dots near the peaks are at the centroid $x=\langle x \rangle$ and
the two outer (blue) rows are distant $\sigma_2$ from $\langle x \rangle$. The dots (black) just left of $\langle x \rangle$ are at $x=\langle x \rangle+\mathcal{S}$. The dashed curves show $|\psi(x)|$ at $t=0$ shifted by $\langle p \rangle t$. \\
(b) Shows the spread $\sigma_2$ (outer (blue) curves), the centroid $\langle x \rangle$ (straight (red) line), and $\langle x \rangle+\mathcal{S}$ (adjacent (black) curve) over a longer time period.}
\end{figure}




\section{Fourth-order moments.} 
For $n=4$, $Y_3=y_3+y_4\,t/m$, and $\textrm{d}_t Y_2 = 2Y_3/m$ shows that $Y_2$, which is quadratic in $t$, takes its minimum value at $t_0 =-my_3/y_4$. Changing to $u=(t-t_0)/m$ in
\begin{equation}\label{eqn:Y0y}
Y_0=y_0+4y_1t/m+6y_2 (t/m)^2+4y_3(t/m)^3+y_4(t/m)^4,	
\end{equation}
that comes from \eref{eqn:Yy}, yields
\begin{equation}\label{eqn:4YT}\eqalign{
	Y_0 = y_4\,u^4 + 6(\Omega_4/y_4)u^2+4(\Lambda_4/y_4^2)u + (\Theta_4/y_4^3),\\
	Y_1= y_4\,u^3 + 3(\Omega_4/y_4)u + (\Lambda_4/y_4^2),\\
	Y_2 = y_4\,u^2 + (\Omega_4/y_4),\\
	Y_3 = y_4\,u,}
\end{equation}
where the equations for $Y_1, Y_2, Y_3$ emerge by successively applying $\textrm{d}_u$ and
\begin{equation}\label{Z4}\eqalign{
	\Omega_4 = y_2\,y_4-y_3^2,\\
	\Lambda_4 = y_1 y_4^2+2y_3^3-3y_2 y_3 y_4,\\
	\Theta_4 = y_0 y_4^3-3y_3^4+6y_2 y_3^2 y_4-4y_1 y_3 y_4^2.}
\end{equation}
It can be easily verified (by applying $\textrm{d}_t$ to $Y_2\,Y_4-Y_3^2$, etc.) that $\Omega_4$, $\Lambda_4$, and $\Theta_4$ are invariant combinations of the moments. From the expressions for the moments in terms of these invariants, we can deduce the general behaviour of time evolution of the moments. 

\begin{figure}\label{fig:Generic4}
\centering
\includegraphics[width=130mm]{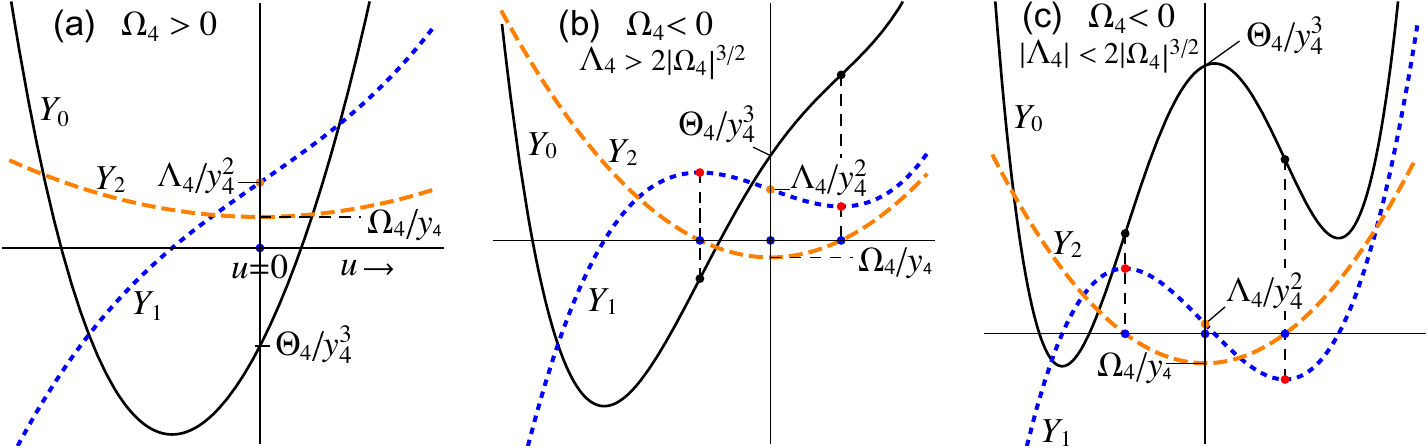}
\caption{Generic forms for the evolution of the fourth moments $Y_0$, $Y_1$, $Y_2$.\\ 
For $\Omega_4<0$, the dashed vertical lines at $u=\pm u_0$ join each inflection of $Y_0$ to the corresponding extremum of $Y_1$. Case (b) with $\Lambda_4 >2|\Omega_4|^{3/2}$ has the minimum of $Y_0$ before $-u_0$. The case, not shown, with $\Lambda_4 <-2|\Omega_4|^{3/2}$, has the minimum after $u_0$.} 
\end{figure}

\subsection{General features of fourth-order moments.} \Eref{eqn:4YT} shows that the sign of $\Omega_4$ will determine whether $Y_2$ has any zeros, and therefore whether $Y_1$ has any extrema, because $\textrm{d}_tY_1=3Y_2/m$. 

\subsubsection*{Case of $\Omega_4>0$.}  $Y_2$ is always positive, so $Y_1$ has no extrema, but has a point of inflection at $u=0$. Also $Y_1-\Lambda_4/y_4^2$ is antisymmetric in $u$ and therefore $Y_1$ will become zero, and $Y_0$ will take its only minimum, at a time with $u<0$ if $\Lambda_4>0$ or with $u>0$ if $\Lambda_4<0$.

\subsubsection*{Case of $\Omega_4\leq0$.} $Y_2$ will be zero at times $\pm u_0$, where $u_0= \sqrt{|\Omega_4|} /y_4$, and take its minimum of $\Omega_4 /y_4$ at $u=0$.

\vspace{1mm}\noindent It follows that $Y_1$ will take:

a maximum of 
$Y_1^+ =(\Lambda_4+2|\Omega_4|^{3/2})/y_4^2$ at $u=-u_0$,

 a decreasing inflection, with $Y_1=\Lambda_4/y_4^2$, at $u=0$, 

 a minimum of 
$Y_1^- =(\Lambda_4-2|\Omega_4|^{3/2})/y_4^2$ at $u=u_0$.

 \noindent The inflections of $Y_0$ occur at the times of the extrema of $Y_1$ and therefore, as time increases from well before $-u_0$, where $Y_0\sim y_4 u^4$, it reaches an inflection at $-u_0$. The sign of the gradient of $Y_0$ at this inflection will determine whether $Y_0$ had a minimum before $-u_0$. 

Thus, if $\Lambda_4 >-2|\Omega_4|^{3/2}$, then $Y_1>0$ at $-u_0$, and $Y_0$ must have passed through a minimum before $-u_0$. 

Similarly, if $\Lambda_4<2|\Omega_4|^{3/2}$, then $Y_1<0$ at $u_0$, and $Y_0$ must have passed through a minimum after $u_0$. 

For two minima we require $2|\Omega_4|^{3/2}>\Lambda_4>-2|\Omega_4|^{3/2}$. Then $Y_0$ must take a maximum between $-u_0$ and $u_0$, at the time when $Y_1=0$. Because  $Y_1$ is decreasing as it passes through its value of $\Lambda_4/y_4^2$ at $u=0$, the maximum of $Y_0$ will be before $u=0$ if $\Lambda_4<0$. [The values of $Y_0$ at the times $\mp u_0$ are $(\Theta_4-5\Omega_4^2\mp 4\sqrt{|\Omega_4|}\,\Lambda_4)/y_4^3$.]

\subsubsection* {In summary, for $\Omega_4<0:$}

If $|\Lambda_4|>2|\Omega_4|^{3/2}$ then $Y_0$ has just one minimum  (it occurs with $u \lessgtr 0$ if $\Lambda_4 \lessgtr 0$), an inflection at $u=u_0$, and $Y_0$ has no local maxima.

If $|\Lambda_4|<2|\Omega_4|^{3/2}$ there is a local minimum with $u<-u_0$, a local maximum with $-u_0<u<u_0$ (and $u \lessgtr 0$ if $\Lambda_4\lessgtr 0$), and a local minimum with $u>u_0$.

\vspace{3mm}
Determination of the precise times for each extremum of $Y_0$ requires the solution of the cubic equation for the zeros of $Y_1$ in \eref{eqn:4YT}; general expressions for the zeros are unwieldy, but for specific cases the zeros can be found numerically. 

\subsection{Multiple extrema is an essentially quantum phenomenon}
More than one critical point in the evolution of $Y_0$ can occur only for $\Omega_4<0$ and it will now be shown that this is excluded for a set of free classical particles, but is possible for a quantum particle.

\subsubsection*{An inequality for $\Omega_4$} 
We use the Schwarz inequality $\langle\hat A^2\rangle \langle\hat B^2\rangle\geq |\langle\hat A\hat B\rangle|^2$ with $\hat A=\frac{1}{2}(\hat P\hat X+\hat X\hat P)$ and $\hat B=\hat P^2$. Applying $ [ \hat P, \hat X ] =-\imath \hbar$ one can show that 
\begin{equation}
\hat A^2=\hat Y_2+\case{1}{4}\hbar^2 \qquad \textrm{and} \qquad \hat A\hat P^2=\hat Y_3+\imath \hbar^2 \hat P^2,
\end{equation}
where $\hat Y_k$ is the symmetrized sum of operators such that $\langle \hat Y_k \rangle=Y_k$.
 The Schwarz inequality becomes
\begin{equation}\label{eqn:ineq}
\Omega_4= Y_2 Y_4-Y_3^2\geq \big(\langle\hat P^2\rangle^2-\case{1}{4}\langle\hat P^4\rangle\big) \hbar^2.
\end{equation}
If the wavefunction is even or odd, $Y_3=0$ and
\begin{equation}
Y_2 \geq \big(\langle\hat P^2\rangle^2/\langle\hat P^4\rangle-\case{1}{4}\big) \hbar^2.
\end{equation} 
The possibility of $\Omega_4<0$ is therefore related to $\langle\hat P^4\rangle/\langle\hat P^2\rangle^2$, the kurtosis of momentum.

We do not use it here, but there is an inequality relating kurtosis to skewness. The Schwarz inequality $\langle \hat X^2\rangle \langle \hat X^4\rangle \geq \langle \hat X^3\rangle^2$ leads to $\langle \hat X^4\rangle / \langle \hat X^2\rangle^2 \geq \langle \hat X^3\rangle^2/ \langle \hat X^2\rangle^3$, and $\langle \hat X^3\rangle/ \langle \hat X^2\rangle^{3/2}$ is a dimensionless measure of the skewness relative to the spread. Similarly,  $\langle \hat P^4\rangle / \langle \hat P^2\rangle^2 \geq \langle \hat P^3\rangle^2/ \langle \hat P^2\rangle^3$, relating the kurtosis of momentum to the skewness of momentum.

\subsubsection*{Negative $\Omega_4$ is non-classical.}
For an ensemble of classical particles, the inequality becomes $\Omega_4\geq0$. [Equation (44) in  Brizuala \cite{BrizGen}.] This follows because the term involving $\hbar^2$ in \eref{eqn:ineq} arises through commutation of the operators and is therefore a purely quantum phenomenon.

Without an example, this derivation does not prove that a state with $\Omega_4<0$ exists; but an example is given in \eref{eqn:psixexp}. Furthermore, this example shows that the lower limit can be approached. (It may be that this limit cannot be actually reached for states such that all fourth moments exist.) The existence of this non-classical behaviour in the fourth-order moments contrasts with that of the second-order moments. For $n=2$, classically $\Omega_2$ must be positive but can be zero, whereas quantum effects require $\Omega_2\geq \frac{1}{4}\hbar^2$. For $n=4$, classically $\Omega_4$ must be positive or zero, but quantum effects allow $\Omega_4$ to be negative. The example given involves an initially real wavefunction.

\subsection{The fourth moments of real wavefunctions.} When $y_1=y_3=0$, it follows that $u_0=0$ and $u=t/m$. Also $\Omega_4=y_2 y_4,\;\Lambda_4=0$, and $\Theta_4=y_0 y_4^3$. The evolution of the moments is
\begin{IEEEeqnarray}{llr}
Y_0\,& = y_0 +6y_2\,u^2+y_4\,u^4, \;\;\,
Y_1 = 3y_2\,u +y_4\,u^3, \;\;\,
Y_2 = y_2 +y_4\,u^2,\;\;\,
Y_3 = y_4\,u.
\end{IEEEeqnarray}
Then $Y_2=0$ for $u^2=-y_2/y_4$. Therefore, if $y_2 > 0$ then $Y_0$ will have no inflections and will have a minimum value of $y_0$ at $u=0$ and no other extrema.

If $y_2 < 0$ then $Y_2=0$ at $u=\pm u_0$ where $u_0=(-y_2/y_4)^{1/2}$, and $Y_0$ will have inflections at those times (and $Y_0=y_0-5y_2^2/y_4$ there). Also $Y_1=0$ at times $u=0$ and $u=\pm\sqrt{3}\,u_0$. Therefore $Y_0$ has a local maximum of $y_0$ at $u=0$ and two equal minima at $u=\pm\sqrt{3}\,u_0$, where it takes the value $Y_0=y_0-9\,y_2^2/y_4$. Furthermore, $Y_0$ passes through the value $y_0$ at the times $u=\pm\sqrt{6}\,u_0$ and at $u=0$.


\subsection{Example with $\Omega_4<0$.} The real wavefunction
\begin{equation}\label{eqn:psixexp}
	\psi(x)=|x|^c\exp(-\case{1}{2}\,|x/a|^b),
\end{equation}
with $a>0$, is singular at $x=0$ (unless b and c are multiples of 2), but all fourth-moments will exist if $b>0$ and $c>3/2$. The details are in Appendix\,A, and the result is that $y_2 = [b\,(2c+1) - 1]\hbar^2/4$. Hence $y_2$ will be negative if $b<1/(2c+1)<1/4$, and therefore $Y_0$ will have two maxima. In this region, the spatial spread of the wavefunction is very large: the second-moment spread $\sigma_2=\langle x^2 \rangle^{1/2}$ approaches its minimum value of approximately $8787a$ at the edge ($b=1/4$ and $c=3/2$) and increases rapidly as $b$ decreases. Because of the slow exponential decrease of $\psi(x)$ for large $x$, the corresponding fourth-moment spread $\sigma_4=\langle x^4 \rangle^{1/4}$ is about twice $\sigma_2$.

For a specific case, with $c=1.51, b=0.24$: $y_2= -0.0088\,\hbar^2,\; \sigma_2\approx 15500\,a, \\ \sigma_4\approx 32000\,a, \;u_0\approx 368$, and the kurtosis $\langle (X/\sigma_2)^4 \rangle\approx18$.\\
The shape of this wavefunction is shown in figure 4.



\begin{figure}\label{fig:ReWfn}
\centering
\includegraphics[width=105mm]{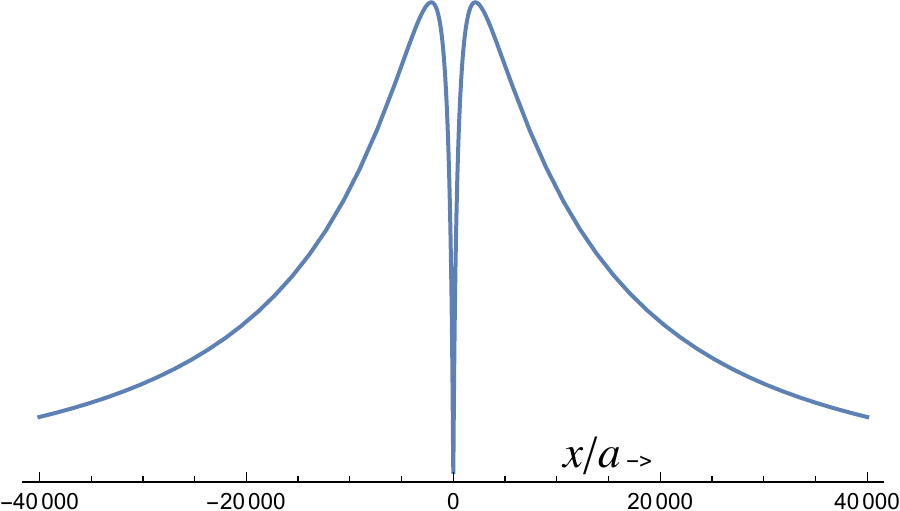}
\caption{The wavefunction $\psi(x)=|x|^c\exp[-\frac{1}{2}|x/a|^b]$\\
with $c=1.51, b=0.24\,$. } 
\end{figure}

\section{Evolution and invariants of moments of any order}\label{section:Z}
The time-derivative $\textrm{d}_t Y_k=(n-k)Y_{k+1}/m$ leads to the higher derivatives
\begin{equation}
\textrm{d}_t^j Y_k=\big[(n-k)!/(n-k-j)!\big]Y_{k+j}/m^j, \;\;\; j=0,...,n-k.
\end{equation}
Taylor's expansion for the evolution is then
\begin{equation}\label{eqn:YkZ}
Y_k=\sum\nolimits_{j=0}^{n-k}[\textrm{d}_t^j Y_{k+j}]_{t=0}\,t^j/j!=
\sum\nolimits_{j=0}^{n-k} (^{n-k}_ {\;\;j})\,y_{k+j}(t/m)^j,
\end{equation}
as in \eref {eqn:Yy}. Similarly, shifting the origin to another time $t_0$ leads to
\begin{equation}\label{eqn:Yt0}
Y_k=\sum\nolimits_{j=0}^{n-k} (^{n-k}_ {\;\;j})\,[Y_{k+j}]_{t=t_0}u^j,
\end{equation}
where $u=(t-t_0)/m$. This holds for any $t_0$, but we will show that taking  $t_0=-m \,y_{n-1}/y_{n}$ leads to an invariant combination of the moments. We define $Z_\ell$ to be  $[Y_{n-\ell}]_{t=t_0}\,y_n^{\ell-1}$, and \eref{eqn:Yt0} becomes
\begin{equation}\label{eqn:YZ}
Y_k=\sum\nolimits_{j=0}^{n-k}\,(^{n-k}_{\;\;j})Z_{n-k-j}u^j/y_n^{n-k-j-1}.
\end{equation}
Furthermore, inserting this time $t_0$, and $k=n-\ell$, into \eref{eqn:YkZ},
\begin{equation}\label{eqn:Zy}
Z_\ell=\sum\nolimits_{j=0}^{\ell}  (^{\ell}_ {j})\,y_{n-\ell+j}(-y_{n-1})^{j}
\,y_{n}^{\ell-j-1},
\end{equation}
The term, $y_{n}^{\ell-1}$ has been included to make $Z_\ell$ a polynomial of degree $\ell \leq n$ in the initial moments for all integer $\ell$. The last two terms in this sum combine, and 
\begin{equation}\label{eqn:Zy2}
Z_\ell=\sum\nolimits_{j=0}^{\ell-2}  (^{\ell-2}_ {\;\;j})\,y_{n-\ell+j}(-y_{n-1})^{j}
\,y_{n}^{\ell-j-1}-(\ell-\!1)(-y_{n-1})^{\ell}.
\end{equation}

Special cases are $Z_{0}=1$, $Z_{1}=0$, and 
\begin{eqnarray}
	Z_{2}&=&-y_{n-1}^2 + y_{n - 2}\, y_{n},\\
	Z_{3}&=&2 y_{n - 1}^3 - 3 y_{n - 2} y_{n - 1} y_{n} + y_{n - 3} y_{n}^2,\nonumber \\ 
Z_{4}&=&-3 y_{n - 1}^4 + 6 y_{n - 2} y_{n - 1}^2 y_{n} - 
  4 y_{n - 3} y_{n - 1} y_{n}^2 + y_{n - 4} y_{n}^3.\nonumber
\end{eqnarray}
Comparing with  \eref{eqn:Z1},   \eref{eqn:Z2}, and \eref{Z4}: $Z_{2}\!=\!\Omega_n,\,Z_{3}\!=\!\Lambda_n, \,Z_{4}\!=\!\Theta_n$.

One can show directly that $\textrm{d}_t\sum_{j=0}^{\ell}(^{\ell}_{j})(-Y_{n-1})^{j}\,Y_{n+j-\ell}\,y_n^{\ell-j-1}=0$ using $\textrm{d}_t Y_k=(n-k)Y_{k+1}/m$, and therefore that every $Z_{\ell}$ is an invariant combination of the moments.\\
(Differentiating $Z_{\ell}$ gives the sum of two terms, one from $\textrm{d}_t (-Y_{n-1})^{j}$ and the other from $\textrm{d}_tY_{n+j-\ell}$. Changing the summation variable in the latter term from $j$ to $j-1$, and using $(^{\ell}_{j})\,j=(^{\;\;\ell}_{j-1})(\ell-j+1)$, shows that the two terms cancel.) This approach, however, does not show clearly why this particular time $t_0$ leads to an invariant.

Another approach uses the Heisenberg picture\,\cite{HP} that allows the operators to change with time in a way that leaves the state unchanged. For a free particle, the operator equations of motion are $\textrm{d}_t \hat X=\hat P/m$ and $\textrm{d}_t \hat P=0$ with the solution $\hat X_t=\hat X_0+\hat P t/m$ and $\hat P$ constant. Then $y_k=\langle \{\hat P^k\hat X_0^{n-k}\}\rangle$ where $ \{\hat P^k\hat X^{n-k}\}$ stands for the symmetrized average. It follows that $Y_k=\langle \{\hat P^k(\hat X_0+\hat Pt/m)^{n-k}\}\rangle$ and $Z_\ell=\langle \{\hat P^{n-\ell}(\hat X_0+\hat Pt_0/m)^\ell\}\rangle$. But $\hat X_0+\hat Pt_0/m=\hat X_0-\hat P y_{n-1}/y_n$, and this evolves to $\hat X_0+\hat Pt/m-\hat P Y_{n-1}/y_n$ = $\hat X_0-\hat P y_{n-1}/y_n$ because $Y_{n-1}= y_{n-1}+y_n t/m$ from \eref {eqn:YkZ}. Therefore  $Z_\ell$ is an invariant combination of the moments.

\section{Conclusions} 

The free evolution of any symmetrized moment of order $n$ can be expressed as a polynomial of degree $n$ in the time, with each coefficient a simple numerical factor multiplying an initial moment. The nature of the evolution, however, is better explored through invariant combinations of the moments. Using a specific origin for the time ($t_0=-m\,y_{n-1}/y_n$), these invariants emerge through the values of the moments at that time-origin. The general features of the evolution of each moment are revealed by its extrema and inflections, and for $n\leq 4$ their sequence is determined by simple combinations of these invariants, as are the precise times and values of these extrema and inflections, although when $n=4$ the solution of a cubic equation is required to obtain some exact values. 

Although the evolution of the moments of a quantum state is in many ways similar to that of a set of classical particles, some local extrema are possible only in the quantum case. An example is $\langle X^4\rangle$, which can have a local maximum only in the quantum case. 

For order $n>4$, a complete analysis of the geometric properties is not possible in terms of simple explicit functions of the initial moments because the zeros of polynomials of degree up to $n-1$ are required. (Partial information can be found from the inflections for $n=5$ and $6$, but this is somewhat obscured by the complexity of the analytic form of the zeros.) Numerical solutions for these zeros yield the geometric properties directly, but only for specific cases.  

\ack I wish to thank Michael J W Hall for valuable comments and suggestions.

\appendix

\section{Further detail on a real wavefunction with negative $\Omega_4$}\label{appendix:C}
For the wavefunction $\psi(x)=|x|^c\exp(-\frac{1}{2}|x/a|^b)$, with $a>0$, all moments up to the fourth order will exist if $b>0$ (so that all required integrals are integrable as $x\to\infty$) and $c>\case{3}{2}$ (so that they are integrable across $x=0$). All required integrals can be exactly integrated in terms of Gamma functions. The second-order moments are:
\begin{IEEEeqnarray}{llr}
\langle X^2\rangle\, & =a^22^{-2/b} \Gamma[b^{-1}(2c+3)]/\Gamma[b^{-1}(2c+1)],	\\
\langle P^2\rangle\, & =\hbar^2a^{-2}2^{-2 + 2/b}[b (2c-1)+1]\,\Gamma[b^{-1}(2c-1)]/\Gamma[b^{-1}(2c+1)],    
\end{IEEEeqnarray}
and the fourth-order results are:
\begin{equation}\eqalign {
y_0 = a^42^{-4/b}\Gamma[b^{-1}(2c+5)]/\Gamma[b^{-1}(2c+1)],\\
y_2 = \hbar^22^{-2}[b (2c+1) - 1],\\
y_4 = \hbar^4a^{-4}2^{-4 + 4/b}f_{b,c}\,   
  \Gamma[b^{-1}(2c-3)]/ \Gamma[b^{-1}(2c+1)],	}
\end{equation}
where $f_{b,c}=(9 + 2 b (c - 3/2 ) (10 + 2 b^2 + 6 b (c - 3/2)))$.

\vspace{2mm}
In relation to the inequality for $\Omega_4$ in \eref{eqn:ineq}, note that $\Omega_4=y_2\, y_4$ and $y_2\to -\case{1}{4}\hbar^2$ as $b\to 0$. Also 
$\langle P^2\rangle^2/\langle P^4\rangle$ is negligibly small (about $2.8\times 10^{-7}$ for $c=1.51$ and $b=0.24$) and approaches zero as $b \to 0$. Thus the inequality approaches saturation in this limit. Also in this limit, however, the wavefunction becomes non-normalizable and $\sigma_2$  becomes infinite. 

It is, of course, possible to cut off the exponential tail of this wavefunction $\psi(x)$ while still achieving a negative $\Omega_4$. As an example we take $c=1.51$ and $b=0.24$, as before, which yielded $\Omega_4\approx -0.0088\,\hbar^2$. Reducing the wavefunction smoothly to zero requires a function that matches $\psi(x)$ and its first and second derivatives, so that $y_4$ exists. If we use the quartic polynomial that matches $\psi(x)$ in this way at $|x|= 6\,\sigma_2$ and such that both it and its first derivative become zero at the cut-off where $|x|= 15\,\sigma_2$, then $\Omega_4\approx -0.0074\,\hbar^2$.

\section{The Wigner function and moments}
The Wigner function\cite{W} $W(x, p)$ is a function in classical phase space that can be generated from the wavefunction. It provides a powerful tool for analysing the free evolution of moments due to two basic properties:\\
1. For free evolution, $W(x, p)$ follows the classical evolution: $W(x,p)_t=W(x+pt, p)_0$.\\ 
2. If $f_S(\hat x, \hat p)$ is the symmetrized operator form of the function $f(x,p)$, its expectation value is given by
 $\langle f_S(\hat x, \hat p)\rangle=\int W (x, p)f (x, p) \, \textrm{d} x \, \textrm{d} p$
 
We will use the notation $\mathcal{W}[f(x,p)]$ to represent $\int W (x, p)f (x, p) \, \textrm{d} x\,\textrm{d} p.$ Then  $y_ k=\mathcal{W}[P^k X^{n - k}]$ and its evolution is
 \begin{equation}\label{eqn:YPX}
\!\!\!\!\!\!\!\!\!\!\!\!\!\!\!\!\!\!\!\!\!\!\!\!Y_k=\mathcal{W}\big[P^k (X + P\,t/m)^{n - k}\big]
 = \sum\nolimits _{j=0}^{n-k}  (^{n - k}_ {\;\;j}) \mathcal{W}\big[P^{k + j} X^{n - k - j} (t/m)^j\big].
 \end{equation}
That is, as in \eref {eqn:Yy},
\begin{equation} \label{eqn:YkA}
Y_k = \sum\nolimits_{j=0}^{n-k}  (^{n - k}_ {\;\;j}) y_ {k + j} (t/m)^j.
\end{equation}

\subsubsection*{Invariants:} We define $Z_\ell$ to be the value of $Y_{n-\ell}\,y_n^{\ell-1}$ at the time $t_0 =-y_{n-1}/y_n$. Then $Z_\ell=\mathcal{W}[P^{n-\ell}(X-P \,y_{n-1}/y_n)^\ell]\,y_n^{\ell-1}$ and this will be invariant because its evolution is $\mathcal{W}[P^{n-\ell}(X+Pt/m-P \,Y_{n-1}/y_n)^{\ell}]\,y_n^{\ell-1}$, and $Y_{n-1}=y_{n-1}+y_{n}t/m$, from \eref{eqn:YkA}. 
Inserting this time $t_0$ into \eref{eqn:YkA},
\begin{equation}
Z_\ell=\sum\nolimits_{j=0}^{\ell}  (^{\ell}_ {j})\,y_{n-\ell+j}(-y_{n-1})^{j}
\,y_{n}^{\ell-j-1},
\end{equation}
as in \eref{eqn:Zy}. This shows that $Z_\ell$ is polynomial in the moments.
Furthermore, $Y_k=\mathcal{W}[P^k(X+P t/m)^{n-k}]=\mathcal{W}[P^k(X-P \,y_{n-1}/y_n+P(t-t_0)/m )^{n-k}]$, and so
\begin{equation}\eqalign{
\!\!\!\!\!\!\!\!\!\!\!\!Y_k=\mathcal{W}\big[\sum\nolimits_{j=0}^{n-k}\,(^{n-k}_{\;\;j})\,P^k(X-P \,y_{n-1}/y_n)^{n-k-j}\,P^j(t-t_0)/m)^j\big] \\
 \hspace{5mm} =\sum\nolimits_{j=0}^{n-k}\,(^{n-k}_{\;\;j})\,Z_{n-k-j}u^j/y_n^{n-k-j-1},}
\end{equation}
where $u=(t-t_0)/m$, as in \eref{eqn:YZ}.


\end{document}